\documentclass[twocolumn,showpacs,preprintnumbers,amsmath,amssymb,showkeys]{revtex4}

\usepackage{graphicx}
\usepackage{dcolumn}
\usepackage{bm}
\usepackage{bezier}

\usepackage{tikz}
\usetikzlibrary{arrows,decorations.pathmorphing,backgrounds,positioning,fit,petri}

\begin{document}

\title{Properties of Open Quantum Walks on $\mathbb{Z}$}
\author{Ilya Sinayskiy and Francesco Petruccione}
\affiliation{National Institute for Theoretical Physics and School of Chemistry and Physics,\\
              University of KwaZulu-Natal, Westville, Durban, South Africa, 4001}%

\date{\today}

\begin{abstract}
A connection between the asymptotic behavior of the open quantum walk and the spectrum of a generalized quantum coins is studied. For the case of simultaneously diagonalizable transition operators an exact expression for probability distribution of the position of the walker for arbitrary number of steps is found. For a large number of steps the probability distribution consist of maximally two ``soliton"-like solution and a certain number of Gaussian distributions. The number of different contributions to the final probability distribution is equal to the number of distinct absolute values in the spectrum of the transition operators. The presence of the zeros in spectrum is an indicator of the ``soliton"-like solutions in the probability distribution. 
\end{abstract}

\pacs{03.65.Yz, 05.40.Fb, 02.50.Ga}
\keywords{Open quantum systems, open quantum walks, transition operator spectrum}
\maketitle

\section{Introduction}
It is well-known that the mathematical concept of classical random walks \cite{pt, rwp} has found wide applications in physics \cite{rwp}, computer science \cite{rwcs}, economics \cite{rwe} and biology \cite{rwb}. 
Basically, the trajectory of a random walk consists of a sequence of random steps on some underlying graph \cite{rwp}. Recently, an extension of the concept of random walk to the quantum domain was performed. Quantum walks can be introduced in a discrete time \cite{aharonov} and in a continuous time \cite{FG} way. For a classical random walk the probability distribution of the position of the walker is given by the transition rates between the vertices of the graph. For the quantum case \cite{kempe} the probability distribution of the walker is defined not only by the transition rates between the nodes of the graph, but also by the dynamics of the internal degrees of freedom of the "walker".  The resulting interference effects is what makes these walks truly quantum.   

Unitary quantum walks found wide application as a tool for the formulation of quantum computing algorithms \cite{qaqrw}. Although, experimental implementation of  any quantum concept is usually difficult due to unavoidable decoherence and dissipation effects \cite{toqs},  realizations of unitary quantum walks have been reported. Implementations with negligible effect of decoherence and dissipation were realized in optical lattices \cite{QWOL}, with photons in waveguide lattices \cite{qwwl}, with trapped ions \cite{qwti} and free single photons in space \cite{qwaw}. 

During the last few years attempts were made to include the effects of decoherence and dissipation in the description of the quantum walks \cite{ken11}. Typically, in these approaches decoherence is treated as an extra modification of the unitary quantum walk scheme, the effect of which needs to be minimized and eliminated.

Recently, a formalism for discrete time open quantum walks, which is exclusively based on the non-unitary dynamics induced by the environment was introduced \cite{longAttal}. The formalism suggested is similar to the formalism of quantum Markov chains \cite{gudder} and rests upon the implementation of appropriate completely positive maps \cite{toqs, kraus}. It was shown that the formalism of the open quantum walks includes the classical random walk and through a physical realization procedure a connection to the unitary quantum walk was established.

Open quantum walks show rich dynamical behavior \cite{longAttal}. The aim of this paper is to analyze in detail the dynamics of the probability distribution for the position of the open quantum walk. Open quantum walks were shown to have a probability distribution which is represented by a sum of Gaussian distributions and singular ``soliton"-like distributions. In particular we will show that there exists a connection  between the number of the Gaussian distributions in the probability distribution and the dimension of the internal degree of freedom of the walker. Furthermore, we will formulate a condition for the appearance of the ``soliton"-like solution in the dynamics of the walker.

The paper is structured as follows. In Section II we briefly revise formalism of open quantum walks. In Section III we study the connection between asymptotic properties of the walk and the structure of the transition operators. In Section IV we conclude.

\section{Open Quantum Walks}

We study a random walk on a set of vertices $\cal{V}$ with oriented edges $\{(i,j)\,;\ i,j\in\cal{V}\}$.  The number of nodes is considered to be finite or countable infinite. The space of states corresponding to the dynamics on the graph specified by the set of nodes $\cal{V}$ will be denoted by  $\cal{K}=\mathbb{C}^\mathcal{V}$. If $\cal{V}$ is an infinite countable set, the space of states $\cal{K}$ will be any separable Hilbert space with an orthonormal basis ${(| i\rangle)}_{i\in\cal{V}}$ indexed by $\cal{V}$. The internal degrees of freedom of the quantum walker, e.g. the spin, angular momenta or $n$-energy levels, will be described by a separable Hilbert space $\cal{H}$ attached to each vertex of the graph. So that, any state of the quantum walker will be described on the direct product of the Hilbert spaces $\cal{H}\otimes \cal{K}$.

To describe the dynamics of the quantum walker for each edge $(i,j)$ we introduce  a bounded operator $B^i_j\in\cal{H}$. This operator describes the change in the internal degree of freedom of the walker due to the "jump" from node $j$ to node $i$ (see Fig.1). By imposing for each $j$ that, 
\begin{equation}\label{eq1}
\sum_i {B^i_j}^\dag B^i_j= I,
\end{equation}
we make sure, that for each node of the graph $j\in\mathcal{V}$ there is a corresponding completely positive map on the positive operators of $\mathcal{H}$:
\begin{equation}
\mathcal{M}_j(\tau)=\sum_i B^i_j \tau {B^i_j}^\dag.
\end{equation}
The operators $B^i_j$ act only on $\mathcal{H}$ and do not perform transitions from node $j$ to node $i$, they can be extended to operators $M^i_j\in\mathcal{H}\otimes\mathcal{K}$ acting on total Hilbert space in the following way
\begin{equation}
M^i_j=B^i_j\otimes | i\rangle\langle j|\,.
\end{equation}
It is clear that, if the condition expressed in Eq. (\ref{eq1}) is satisfied, then $\sum_{i,j} {M^i_j}^\dag M^i_j=1$. This condition defines a completely positive map for  density matrices on $\mathcal{H}\otimes\mathcal{K}$, i.e.,
\begin{equation}\label{DQRW}
\mathcal{M}(\rho)=\sum_i\sum_j M^i_j\,\rho\, {M^i_j}^\dag.
\end{equation}
The above  map defines the discrete time \textit{Open Quantum Walk} (OQW).
It is easy to see that for an arbitrary initial state the density matrix $\sum_{i,j} \rho_{i,j}\otimes| i\rangle\langle j|$ will take a diagonal form after just one step of the open quantum  random walk Eq. (\ref{DQRW}). Hence, in the following,  we will assume that the initial state of the system has the form
\begin{equation}\label{rho}
\rho=\sum_i \rho_i\otimes | i\rangle\langle i|,
\end{equation}
with
\begin{equation}
\sum_i \mathrm{Tr}(\rho_i)=1.
\end{equation}
It is straightforward to give an explicit iteration formula for the OQW from step $n$ to step $n+1$ 
\begin{equation}\label{rho}
\rho^{[n+1]}=\mathcal{M}(\rho^{[n]})=\sum_i \rho_i^{[n+1]}\otimes | i\rangle\langle i|,
\end{equation}
where
\begin{equation}
\rho_i^{[n+1]}=\sum_j B^i_j \rho_j^{[n]}{B^i_j}^\dag.
\end{equation}
The above iteration formula gives a clear physical meaning to the mapping that we introduced: the state of the system on  site $i$ is determined by the conditional shift from all connected sites $j$, which are defined by the explicit form of the generalized quantum coin operators $B_j^i$. Also, one can see that $\mathrm{Tr}[\rho^{[n+1]}]=\sum_i\mathrm{Tr}[\rho_{i}^{[n+1]}]=1$.

\begin{figure}
\includegraphics[width= .9\linewidth]{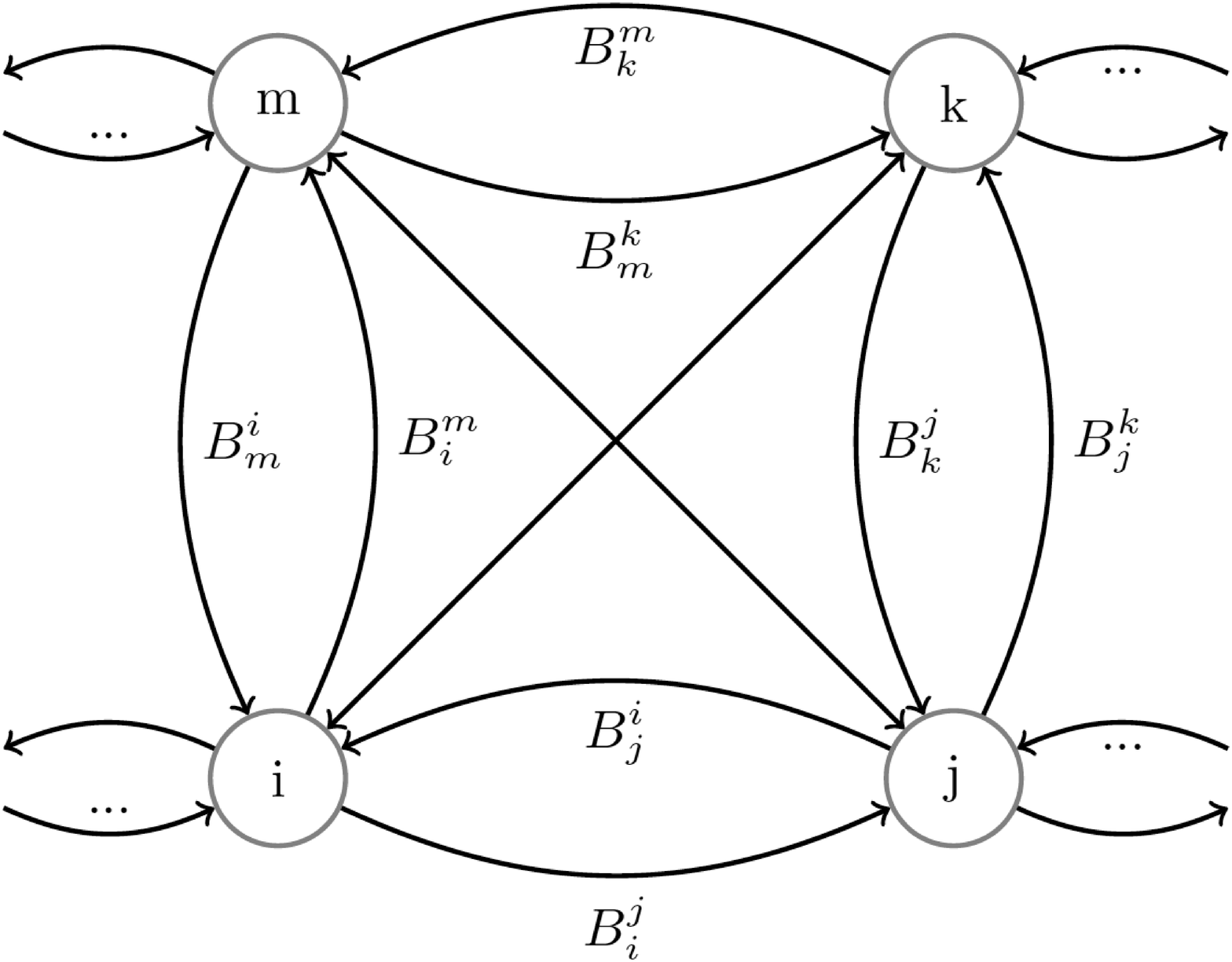}
\caption{Schematic illustration of the formalism of the Open Quantum Random Walk: The walk is realized on a graph with a set of vertices denoted by $i,j,k,m\in\cal{V}$.  The operators $B_i^j$ decribes transitions in the internal degree of freedom of the ``walker" jumping from node $(i)$ to node $(j)$.}
\end{figure}

\section{Properties of transition operators}

Open quantum walks, even on the line show rich dynamical behavior \cite{longAttal}. In this paper we concentrate on the particular case of OQWs on $\mathbb{Z}$ with transition between neighboring nodes (see Fig. 2a). In this case, the general expression for the open quantum walk on the graph reads as,
\begin{equation}\label{rhoz}
\rho^{[n+1]}=\mathcal{M}(\rho^{[n]})=\sum_i \rho_i^{[n+1]}\otimes | i\rangle\langle i|,
\end{equation}
where
\begin{equation}
\rho_i^{[n+1]}=B^i_{i+1} \rho_{i+1}^{[n]}{B^{i\dag}_{i+1}}+B^i_{i-1} \rho_{i-1}^{[n]}{B^{i\dag}_{i-1}}.
\end{equation}
In this paper we will analyze a homogenous open quantum walk (which implies that $\forall i,B_i^{i+1}\equiv B$ and $B_i^{i-1}\equiv C$) with simultaneously diagonalizable generalized coin operators, i.e. $[B,C]=0$.

As the first example of the open walk on $\mathbb{Z}$ let us consider an open quantum walk with a two dimensional Hilbert space for the generalized quantum coins, i.e., $B,C\in \mathbb{C}^2$.
Specifically, we consider the transition operators $B$ and $C$ defined as
\begin{equation}\label{BCZ}
B=\left(\begin{array}{cc} 1 & 0 \\ 0 & \cos{\theta}
\end{array}\right),\quad C=\left(\begin{array}{cc} 0 & 0 \\ 0 & \sin{\theta}\end{array}\right).
\end{equation}
It is clear that the above $B$ and $C$ satisfy the normalization condition, i.e., $B^\dag B+C^\dag C=I$. If initially the waker is in the node $0$,
\begin{equation}
\label{initcond1}
\rho_{0}=\left(\begin{array}{cc} p & z \\ z^* & q \end{array}\right)\otimes  | 0\rangle\langle 0|,
\end{equation}
then the probability distribution of the position of the walker after $n$ steps has the following form,
\begin{eqnarray}
P_k^{[n]}&=&\mathrm{Tr}[\langle k|\rho^{[n]}|k\rangle]=\\\nonumber
& &p\delta_{n,k}+q\left(\begin{array}{c} n \\ \frac{n+k}{2} \end{array}\right)(\sin^2 \theta)^{(n-k)/2}(\cos^2 \theta)^{(n+k)/2},
\end{eqnarray}
where the nonzero values of the probability distribution corresponds to $|k|\leq n$ and $n\pm k$ should be even. The last condition means that after an odd number of steps only odd nodes can be populated and after even number of steps only even ones.

Two distinct behaviors can be identified. First is a trapped state which propagates in the positive direction and second a binomially distributed part, which for a large number of steps becomes Gaussian. One can calculate the mean and the variance of the binomial distribution,
\begin{equation}
\mathbb{E}(x_n)=\cos{(2\theta)}n,\quad \mathrm{Var}(x_n)=\sin{(2\theta)}\sqrt{n},
\end{equation}
where $x_n$ denotes the binomial distribution part of the position of the walker after $n$-steps.

\begin{figure}
\includegraphics[width= .9\linewidth]{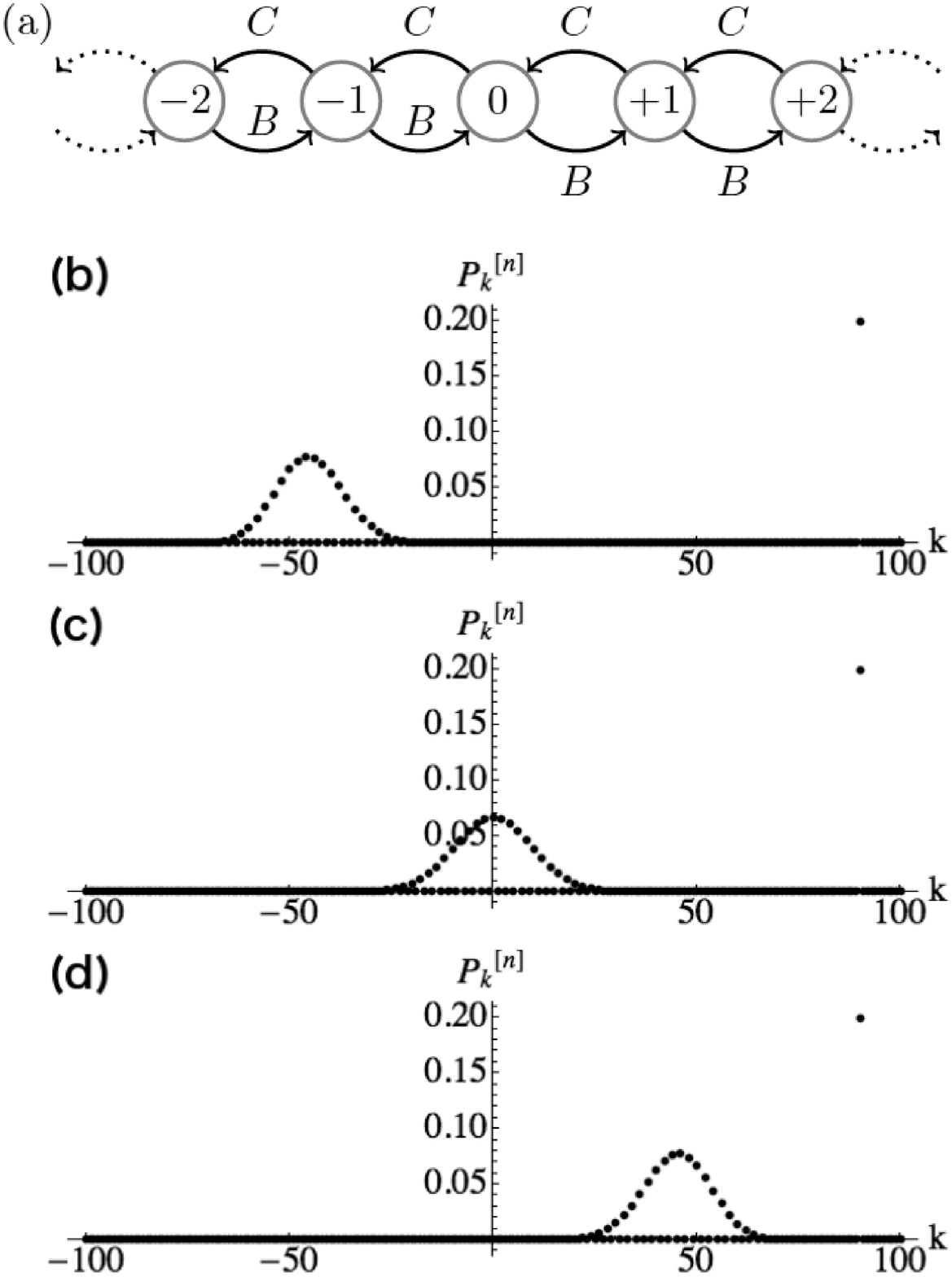}
\caption{OQRW on $\mathbb{Z}$. (a) A schematic representation of the  OQRW on $\mathbb{Z}$: all transitions to the right are induced by the operator $B_i^{i+1}\equiv B$, while all transitions to the left are induced by the operator $B_i^{i-1}\equiv C$; Figures (b)-(d) show the occupation probability distribution for the ``walker" after 90 steps with the initial state given by Eq.(\ref{initcond1}) with values $p=0.2, q=0.8, z=0.1$
and transition operators given by Eq. (\ref{BCZ}) for parameter $\theta$ equals to $\pi/3$, $\pi/4$ and $\pi/6$, respectively.}
\label{fig:DQRWZ1}
\end{figure}

We consider now the more general case of an open quantum walk with a $n$-dimensional Hilbert space for the transition operators. Again, we require the generalized quantum coin operators to be simultaneously diagonalizable. This implies that the generalized coins $B$ and $C$ can be written in the form,
\begin{equation}
B=\sum_{i=1}^n\lambda_i | b_i\rangle\langle b_i|, \quad C=\sum_{i=1}^n\phi_i | b_i\rangle\langle b_i|.
\label{expan}
\end{equation}
The normalization condition, $B^\dag B+C^\dag C=I$ implies that, $\forall i, |\lambda_i|^2+|\phi_i|^2=1$.
In this case the probability distribution of the position of the walker after n steps has the form,
\begin{eqnarray}
P_k^{[n]}&=&\mathrm{Tr}[\langle k|\rho^{[n]}|k\rangle]=\\\nonumber
& &\sum_{i=1}^np_i\left(\begin{array}{c} n \\ \frac{n+k}{2} \end{array}\right)(|\phi_i|^2)^{(n-k)/2}(|\lambda_i|^2)^{(n+k)/2},
\end{eqnarray}
where the coefficients $p_i$ are the populations of the corresponding levels of the initial state in the basis $|b_i\rangle$, i.e.,
\begin{equation}
p_i=\langle b_i,0|\rho_0|b_i,0\rangle.
\end{equation}
The mean and variance of each binomial distribution can be calculated explicitly,
\begin{eqnarray}
\mathbb{E}_i(x_n)&=&n\left(|\lambda_i|^2-|\phi_i|^2\right),\\\nonumber \mathrm{Var}_i(x_n)&=&2|\lambda_i||\phi_i|\sqrt{n}.
\end{eqnarray}

The explicit knowledge of the probability distribution allows to describe the asymptotic behavior of the open quantum walk on $\mathbb{Z}$. It is clear that for generalized quantum coins $B$ and $C$ that  are simultaneously diagonalizable, there are two cases. First, if all coefficients $\lambda_i$ are different and have value between $0$ and $1$, i.e.,
\begin{equation}
\forall i\neq j, |\lambda_i|\neq |\lambda_j|,\quad 0<|\lambda_j|<1,
\end{equation}
then for a large number of steps the probability distribution consists of $n$ Gaussian distributions, where $n$ is the dimension of the Hilbert space of the generalized quantum coins.
If some of the eigenvalues of the generalized quantum coins are degenerate, then the number of Gaussians in the distributions is given by the number of distinct eigenvalues of the operators $B$ and $C$.
Second, if one or more of eigenvalues $|\lambda_i|$ are equal to $0$ or to $1$, then for a large number of steps the probability distribution contains ``soliton"-like distributions in the probability distribution. If one of the $\lambda_i\equiv 0$ (which means that the corresponding $|\phi_i|\equiv1$), then the probability distribution contain a ``soliton"-like solution propagating ballistically in the negative direction; if one of the  $|\lambda_i|\equiv 1$ then a ``soliton"-like solution propagates in the positive direction (see Fig2(b)-(d)).

It is clear, that in the case considered here ($[B,C]=0$) the probability distribution can have $m$ components,i.e., Gaussians and solitons, where $m$ is number of distinct absolute values of eigenvalues of transition operators $B$ and $C$. This includes maximally two "soliton"-like solutions propagating in opposite directions.

\section{Conclusions}

In conclusion, we gave a brief revision of the open quantum walk formalism and studied a connection between the asymptotic behavior of the open quantum walk and the spectrum of the generalized quantum coins. In particular, we consider the case of the open quantum walk on the line with simultaneously diagonalizable transition operators $B$ and $C$. We have found exact expressions for the probability distribution of the position of the walker for arbitrary number of steps. We have shown that for a large number of steps the probability distribution consists of maximally two ``soliton"-like solution and a certain number of Gaussian distributions. We also provide explicit expression for the mean and the variance for each of the Gaussians as the function of the spectrum of the generalized quantum coin operators. We found that a number of different contributions to the final probability distribution is equal to the number of distinct absolute values in the spectrum of the transition operators. We have shown that the presence of zeros in the spectrum is an indicator of the ``soliton"-like solutions in probability distribution. In future, we plan to extend the present analysis to the case of the generic transition operators.   


\begin{acknowledgments}
This work is based upon research supported by the South African Research Chair Initiative of the Department of Science and Technology and National Research Foundation.
\end{acknowledgments}

\end{document}